\documentstyle[prd,aps]{revtex}

\begin{document}
\tightenlines
\draft
\preprint{UOM/NPh/HQP/00-2}
\title{Spectators effect in inclusive beauty decays\footnote{To appear
in the Proceedings of the Sixth Workshop on High Energy Physics
Phenomenology held at the Institute of Mathematical Sciences, Chennai,
India during Jan. 3-15, 2000.}}
\author{S. Arunagiri}
\address{Department of Nuclear Physics, University of Madras,\\
Guindy Campus, Chennai 600 025, Tamil Nadu, INDIA}
\maketitle
\begin{abstract}
I review the role of the spectator quarks effect in the inclusive
beauty decays. The evaluation of the expectation values of four-quark
operators between hadronic states and its consequences
are discussed.
\end{abstract}
\vskip1cm
Inclusive decays of heavy hadrons are described by the heavy quark
expansion (HQE), an expansion in the inverse powers of the
heavy quark mass ($m$) based on the operator product expansion (OPE)
in QCD and the heavy quark effective theory (HQET) assuming
quark-hadron duality \cite{neubert}. The leading order hadronic
decay rate, proportional to $m^5$, is that of the free heavy quark.
Corrections appear at $O(1/m^2)$ and beyond. They are due to
the heavy quark motion inside the hadron and the chromomagnetic
interaction at $O(1/m^2)$ and the spectator quarks processes
at $O(1/m^3)$. The decay rate at order two in $1/m$ splits up
into the mesonic one on the one hand and the baryonic on the other.
This is because of the vanishing chromomagnetic interaction
in the baryons with an exception of $\Omega_Q$. Among the predictions
of the HQE for the inclusive properties which are confronted
by the experimental values like the lifetime of $\Lambda_b$,
semileptonic branching ratio of $B$ and the charm counting in the final
state \cite{hy}, we address the ratio $\tau(\Lambda_b)/\tau(B^0)$
which is 0.9 by theory but 0.79 from experiment),
the specators effect in charmless semileptonic decay of
$\Lambda_b$ on $Br(b \rightarrow X_u l\nu_l)$ and
the validity of the assumption of quark-hadron duality. 

In view of the discrepancy of the theoretical prediction with
the experimental one for $\tau(\Lambda_b)$, it is necessary to
accomodate the contribution coming from the third order term in the
HQE:
\begin{equation}
C(\mu)<H|(\bar b \Gamma q)(\bar q \Gamma b)|H>
\end{equation}
where the Wilson coefficient, $C(\mu)$, describes the spectator quarks
processes: in the decay $Q(q) \rightarrow Q^\prime q_1 q_2 (q)$,
if either $q_1$ or $q_2$ is the same as $q$, then both of them interfere
destructively; if $q_1$ or $q_2$ is the antiquark of $q$, then
they weakly annihilate; and the other one is the $W$-scattering:
$Qq_{1(2)}W \rightarrow Q^\prime q_{2(1)}$. These processes are found
to enhance the decay rate of $\Lambda_Q$. On the other hand, the
central issue in the systematic incorporation of the spectators
effect is the evaluation of the expectation values of
the four-quark operators ($EV_{FQO}$). Traditionally,
for mesons, the $EV_{FQO}$ is obtained, with the vacuum
saturation approximation, in terms of the leptonic decay
constant of the hadron, $f_H$; on the other hand, for baryons,
the valence quark model is employed. This procedure and other
methods \cite{rosner,neubert1,colangelo} found that the FQO
do not account for the discrepancy. However, we have shown in our
recent works \cite{arun1,arun2} that the FQO accounts for
the difference in the lifetimes of $\Lambda_b$ and B.

The $EV_{FQO}$ between hadronic states is related to the form
factor characterising the light quark scattering off the heavy
quark inside the hadron \cite{pirjol}:
\begin{equation}
{1 \over {2M_H}}<H|(\bar b \Gamma q)(\bar q \Gamma b)|H> =
|\Psi(0)|^2 = \int {d^3 \over {(2\pi)^3}} F(q^2)
\end{equation}
In \cite{arun1}, representing the form factor by
$e^{-q^2/4\beta^2}$, the wave function density is obtained as
\begin{equation}
|\Psi(0)|^2 = {\beta^3 \over {4\pi^{3/2}}}
\end{equation}
where $\beta$ is determined by solving the Schr\"odinger
equation in Variational procedure for the wave function
${\beta^{3/2} \over {\pi^{3/4}}}e^{-\beta^2 r^2/2}$
with the potential $V(r)_{meson} = a/r+br+c$ and
$V(r)_{baryon} = a/r+br+\beta r^2+c$. In this description,
the baryon is considered as a two body system of a heavy quark-diquark.
The $\beta$'s for the hadrons are:
$\beta_{B^-} = 0.4, \beta_{B^s} = 0.44$ and $\beta_{\Lambda_b}$ = 0.72,
all in $GeV$ units. Using these values for the wave function density,
the ratio of lifetimes of $\Lambda_b$ and $B$ is found to be 0.79.

If one assumes that the HQE is an asymptotic expansion, then
the expansion for the decay rate can safely be considered
as converging at $O(1/m^3)$. Recently, Voloshin \cite{volo} has
analysed the relations between the inclusive
decay rates of the charmed and beauty baryons triplet
($\Lambda_Q, \Xi_Q$) \footnote{Voloshin did not
consider that the expansion converges at the third order which is
consequencial.}.
The relations depend only on the HQE
and on the flavour symmetry under $SU(3)_f$. In this procedure, the
EV$_{FQO}$ between baryon states is obtained using the
differences in the total decay rates. In \cite{arun2},
strongly assuming that the HQE converges at $O(1/m^3)$,
we extended the Voloshin analysis to $SU(3)_f$ triplet
of the $B$ mesons, $B^-$, $B^0$ and $B^0_s$.
Their total decay rate splits up due to their light 
quark flavour dependence at the third order
in the HQE. The differences in the decay rates of the triplet,
are related to the third order terms
in $1/m$ by
\begin{eqnarray}
d\Gamma_{B^0-B^-} &=& -\Gamma^\prime_0 (1-x)^2
\left\{Z_1{1 \over 3}(c_0+6)+(c_0+2)\right\}
\left<O_6\right>_{B^0-B^-}\\
d\Gamma_{B^0_s-B^-} &=& -\Gamma^\prime_0 (1-x)^2
\left\{Z_2{1 \over 3}(c_0+6)+(c_0+2)\right\}
\left<O_6\right>_{B^0_s-B^-}\\
d\Gamma_{B^0_s-B^0} &=& -\Gamma^\prime_0 (1-x)^2
\left\{(Z_1-Z_2){1 \over 3}(c_0+6)\right\}
\left<O_6\right>_{B^0_s-B^0}
\end{eqnarray}
On the other hand, for the triplet baryons,
$\Lambda_b$, $\Xi^-$ and $\Xi^0$, with
$\tau(\Lambda_b)$ $<$ $\tau(\Xi^0) \approx \tau(\Xi^-)$,
we have the relation between
the difference in the total decay rates and the terms of
$O(1/m^3)$ in the HQE, as
\begin{equation}
d\Gamma_{\Lambda_b-\Xi^0} = {3 \over 8}\Gamma_0^\prime (c_{00}-2)
\left<O_6\right>_{\Lambda_b-\Xi^0}
\end{equation}

For the decay rates $\Gamma(B^-)$ = 0.617 $ps^{-1}$,     
$\Gamma(B^0)$ = 0.637 $ps^{-1}$ and $\Gamma(B^0_s)$ = 0.645 $ps^{-1}$,
the EV$_{FQO}$ are obtained for $B$ meson, as an average, from
Eqs. (4-6): $\left<O_6\right>_B = 8.08 \times 10^{-3} GeV^3$.
The EV$_{FQO}$ for the baryon
$\left<O_6\right>_{\Lambda_b-\Xi^0} = 3.072 \times 10^{-2} GeV^3$,
where we have used the decay rates corresponding to the lifetimes
1.24 $ps$ and 1.39 $ps$ of $\Lambda_b$ and $\Xi^0$ respectively.
The EV$_{FQO}$  for baryon is about 3.8 times larger than that of B.
For these values $\tau(\Lambda_b)/\tau(B) = 0.78$. Using the
experimental value of $\tau(B^-)$ = 1.55 $ps$ alongwith the
above theoretical value, the lifetime of $\Lambda_b$ turns out to be
$\tau(\Lambda_b)$ = 1.20 $ps$. 

We now turn up to the spectator quarks effect in $\Lambda_b \rightarrow
\Lambda_u l \nu_l$ \cite{arun3}, in view of the ALEPH
measurement\cite{aleph} of $Br(b \rightarrow X_u l \nu_l)$.
When $b$ decays into $u l \nu_l$,
the final state $u$ quark constructively interferes with the $u$
quark in the initial state. This effect increases the decay rate
leading to the ratio, using the $EV_{FQO}$ for baryons obtained above, 
\begin{equation}
{\Gamma(\Lambda_b \rightarrow X_u l\nu_l) \over
{\Gamma(b \rightarrow X_u l\nu_l)}} = 1.34
\end{equation}
The $b$-baryon contributes about 10\% to
$Br(b\rightarrow X_u l \nu_l)$. The above estimate will have effect on the
branching ratio considerably if there is no compensation from
elsewhere. The esitmate above will increase if the spectators
effect from $\Xi^0$. It seems that any compensation is absent to offset
the above estimate plus that one from $\Xi^0$. This will, though
modestly, effect the value of the CKM matrix element $|V_{ub}|$.

Concerning quark-hadron duality in the heavy hadron decays,
we make inference that follows the results obtained above.
The agreement found between theory and experiment for
$\tau(\Lambda_b)$, besides consistency in the $B$ mesons case,
clearly signals that quark-hadron duality holds good in the HQE.
In the previous case, Eqs. (2-3), by the choice of the form factor
representation, we obtained $\tau(\Lambda_b)/\tau(B)$,
whereas in the latter it is the very assumption that
the HQE converges at $O(1/m^3)$ which leads to the prediction
for the ratio. The validity of the assumption that we made needs
to be verified \cite{arun4}. In the recent lattice study\cite{sach},
the authors stated that the role of the FQO is significant to explain
$\tau(\Lambda_b)$. We hope that their claim will throw light.

In conclusion, we make a note of warning.
The evaluation of the $EV_{FQO}$ is model dependent
one in the first case\cite{arun1} and is subject to the validity
of the assumption on the convergence of the HQE in the latter
one\cite{arun2}. The intriguing point is that $|\Psi(0)|^2$ for meson
is smaller that the esitmate in terms of the leptonic decay constant.
The ratio $|\Psi(0)|^2_{\Lambda_b}/|\Psi(0)|^2_B$ is larger
than expected.  
\acknowledgements
{The author is grateful to Prof. H. Yamamoto, Prof. Rahul Sinha,
Dr Anjan Giri, Dr Rukmani Mahanta and Mr K. R. S. Balaji for
useful discussions. I acknowledge the encouragement being shown by
Prof. P. R. Subramanian and Dr. D. Caleb Chanthi Raj.

\end{document}